\documentclass[10pt,conference]{IEEEtran}

\usepackage{flushend}
\usepackage{multicol}
\usepackage{lipsum}
\usepackage{subcaption}
\usepackage{multirow}

\usepackage{lipsum}
\usepackage{cite}
\usepackage{amsmath,amssymb,amsfonts}
\usepackage{algorithmic}
\usepackage{graphicx}
\usepackage{textcomp}
\usepackage{xcolor}
\def\BibTeX{{\rm B\kern-.05em{\sc i\kern-.025em b}\kern-.08em
    T\kern-.1667em\lower.7ex\hbox{E}\kern-.125emX}}

\usepackage{url}
\usepackage{hyperref}
\hypersetup{
    colorlinks=true,
    citecolor=black,linkcolor=black,
    filecolor=black, urlcolor=cyan,
}

\usepackage{tcolorbox}
\usepackage{dblfloatfix}
\usepackage{booktabs}

\newtcolorbox{boxH}{
    colback = white!90!gray, 
    colframe = black, 
    boxrule = 0pt, 
    leftrule = 3pt
}

\newcommand{\eg}{\emph{e.g.,}}
\newcommand{\ie}{\emph{i.e.,}}
\newcommand{\etal}{\emph{et al.}}

\usepackage{listings}
\definecolor{codegreen}{rgb}{0,0.6,0}
\definecolor{codegray}{rgb}{0.5,0.5,0.5}
\definecolor{codepurple}{rgb}{0.58,0,0.82}
\definecolor{backcolour}{rgb}{0.95,0.95,0.92}
\lstdefinestyle{mystyle}{
    commentstyle=\color{codegreen},
    keywordstyle=\color{magenta},
    numberstyle=\tiny\color{codegray},
    stringstyle=\color{codepurple},
    basicstyle=\ttfamily\footnotesize,
    breakatwhitespace=false,         
    breaklines=true,                 
    captionpos=b,                    
    keepspaces=true,                 
    numbers=left,                    
    numbersep=5pt,                  
    showspaces=false,                
    showstringspaces=false,
    showtabs=false,                  
    tabsize=2
}

\begin{document}

\title{TestMigrationsInPy: A Dataset of Test Migrations from Unittest to Pytest}


\author{
\IEEEauthorblockN{Altino Alves, Andre Hora}
\IEEEauthorblockA{
\textit{Department of Computer Science, UFMG}\\
Belo Horizonte, Brazil \\
{\{altinojunior, andrehora\}}@dcc.ufmg.br}
}

\maketitle

\begin{abstract}
Unittest and pytest are the most popular testing frameworks in Python.
Overall, pytest provides some advantages, including simpler assertion, reuse of fixtures, and interoperability.
Due to such benefits, multiple projects in the Python ecosystem have migrated from unittest to pytest.
To facilitate the migration, pytest can also run unittest tests, thus, the migration can happen gradually over time.
However, the migration can be time-consuming and take a long time to conclude.
In this context, projects would benefit from automated solutions to support the migration process.
In this paper, we propose TestMigrationsInPy, a dataset of test migrations from unittest to pytest.
TestMigrationsInPy contains 923 real-world migrations performed by developers.
Future research proposing novel solutions to migrate frameworks in Python can rely on TestMigrationsInPy as a ground truth.
Moreover, as TestMigrationsInPy includes information about the migration type (\eg~changes in assertions or fixtures), our dataset enables novel solutions to be verified effectively, for instance, from simpler assertion migrations to more complex fixture migrations.
TestMigrationsInPy is publicly available at: \url{https://github.com/altinoalvesjunior/TestMigrationsInPy}.
\end{abstract}




\section{Introduction}

Unittest~\cite{unittest} and pytest~\cite{pytest} are the most popular testing frameworks in Python~\cite{jetbrains_python, barbosa2022}.
Unittest belongs to the Python standard library and pytest is a third-party testing framework.
In pytest, tests can be regular functions, while unittest tests are contained in classes that inherit from \texttt{TestCase}.
Consequently, pytest tests tend to be less verbose than unittest ones.
Another difference is the assertions: unittest provides \texttt{self.assert*} methods, while pytest allows developers to use the regular \texttt{assert} statement for verifying expectations and values.
Overall, pytest provides some advantages compared to unittest, including simpler assertion, reuse of fixtures, and interoperability~\cite{pytest, barbosa2022}.

\begin{figure}[t]
     \centering
         \includegraphics[width=0.48\textwidth]{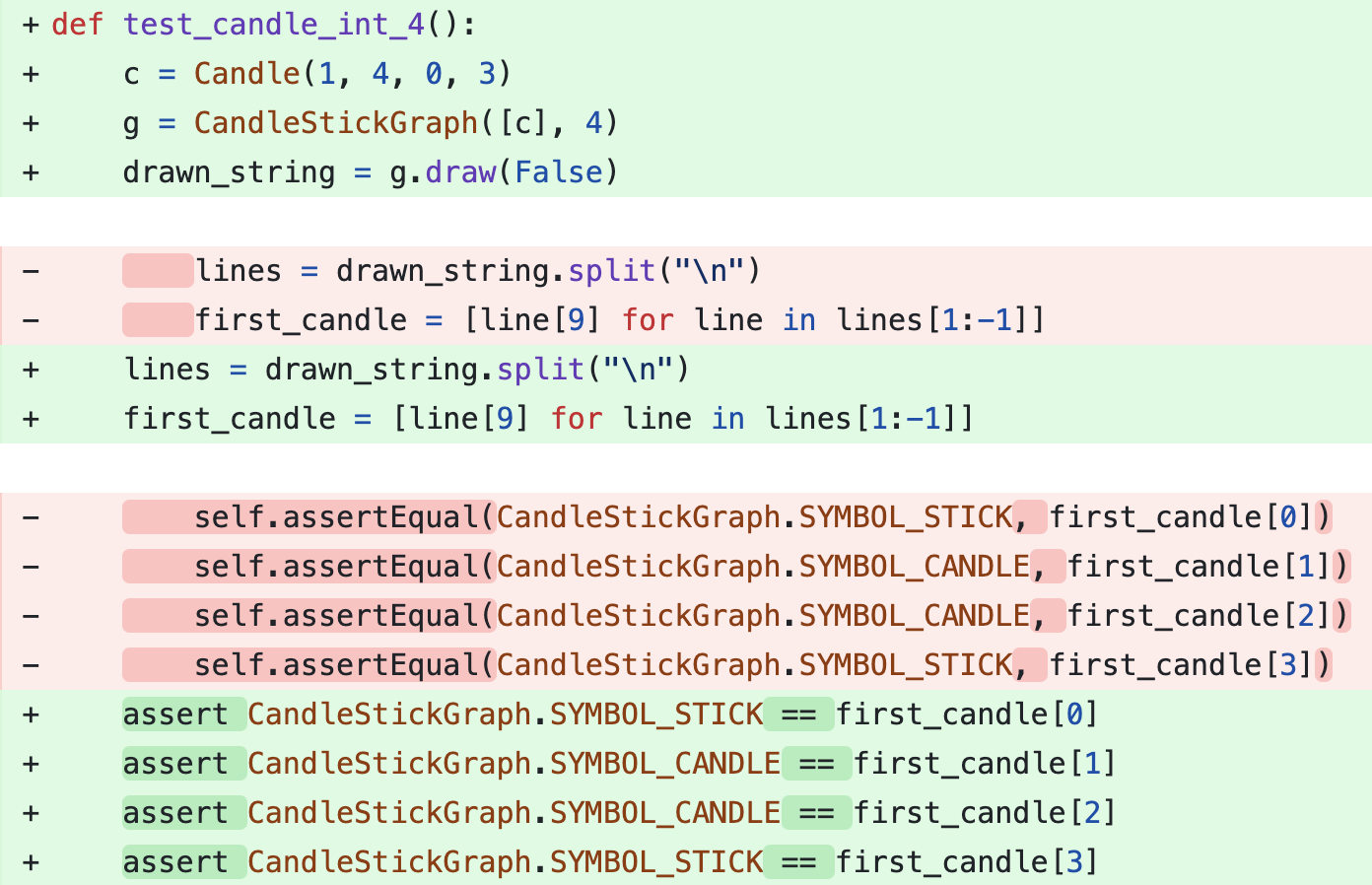}
         \caption{Migration from unittest to pytest (Termgraph).}
        \label{fig:test_candle_int_4}
\end{figure}

Due to such benefits, multiple projects in the Python ecosystem have migrated from unittest to pytest~\cite{barbosa2022}.
A prior study found that 27\% of top-100 most popular Python projects migrated or were migrating to pytest~\cite{barbosa2022}.
Figure~\ref{fig:test_candle_int_4} presents a migration from unittest to pytest in project Termgraph.\footnote{\url{https://github.com/sgeisler/termgraph/commit/d5665248b7d596cabe0a5}}
The unittest \texttt{self.assertEqual} methods (red) are replaced by \texttt{assert} statements in pytest (green).
As the pytest test becomes a function, the inheritance is not needed anymore.

A migration that involves setups and teardowns may be harder to achieve because there is no direct mapping between unittest and pytest.
For example, in project pyvim, the unittest \texttt{setUp} method is split into four pytest fixture functions.\footnote{\url{https://github.com/prompt-toolkit/pyvim/commit/7e1c7bfb505cefba468}}
To facilitate the migration process, pytest can also run unittest tests, meaning that Python test suites can have both testing frameworks simultaneously. 
Consequently, the migration can happen gradually over time.
However, there are also drawbacks: the migration process can be time-consuming and take a long time~\cite{barbosa2022}.
In this context, projects would benefit from automated solutions to support the migration process.

In this paper, we propose TestMigrationsInPy, a dataset of test migrations from unittest to pytest.
TestMigrationsInPy contains 923 real-world migrations performed by developers.
Our dataset construction has two major steps: (1) we automatically detect commits with migrations from unittest to pytest, and (2) we manually inspect the migration commits to identify isolated migrations.
We envision the following usages for TestMigrationsInPy.
First, future research proposing novel solutions to migrate frameworks in Python can rely on TestMigrationsInPy as a ground truth.
Second, as TestMigrationsInPy also includes information about the migration type (\eg~changes in assertions or fixtures), our dataset enables novel solutions to be verified effectively, from simpler assertion migrations to more complex fixture migrations. 

\smallskip

\noindent\textbf{Originality:}
To our knowledge, this is the first dataset in the context of testing framework migration.
Particularly, we focus on a highly relevant migration in the Python ecosystem: unittest to pytest ~\cite{barbosa2022}.

\smallskip

\noindent\textbf{Data Availability:}
TestMigrationsInPy is publicly available at: \url{https://github.com/altinoalvesjunior/TestMigrationsInPy}.

\section{Dataset Construction}

Our dataset construction has two major steps.
First, we automatically detect commits with migrations from unittest to pytest (Section~\ref{subsec:step1}).
Second, we manually inspected the migration commits to identify isolated migrations, that is, migrations that simply replace unittest by pytest (Section~\ref{subsec:step2}).

\subsection{Detecting Migrations from Unittest to Pytest}
\label{subsec:step1}

We rely on the migration detection tool proposed by Barbosa and Hora~\cite{barbosa2022} to detect systems that migrated from unittest to pytest.
We adopt this tool because it has a precision and recall of 100\% in detecting migrations.
The tool classifies the system as \emph{migrated} (or as \emph{is migrating}) when it has at least one migration commit.
A \emph{migration commit} is a commit that explicitly migrates code from unittest to pytest.
To detect migration commits, it assesses the version history of the system with the support of PyDriller~\cite{spadini2018pydriller}.
For a given system, it iterates on its commits and analyzes the \emph{removed} and \emph{added} lines of code per commit.
A commit is a \emph{migration commit} if at least one of the following migration types is true~\cite{barbosa2022}:

\begin{enumerate}
    
    \item \textbf{Assert migration}: the commit removes unittest \texttt{self.assert*} and adds pytest \texttt{assert} statements.
    
    \item \textbf{Fixture migration}: the commit removes unittest setups/teardowns, and adds pytest fixtures.
    
    \item \textbf{Import migration}: the commit removes \texttt{import unittest} and adds \texttt{import pytest}.
    
    \item \textbf{Skip migration}: the commit removes unittest test skips and adds pytest test skips.
    
    \item \textbf{Expected failure migration}: the commit removes unittest expected failure and adds pytest expected failure.
    
\end{enumerate}

\subsection{Detecting Isolated Migrations}
\label{subsec:step2}

The dataset is built based on the manual analysis of migration commits collected in the previous step. 
It is important to notice that a migration commit may have one or more migrations from unittest to pytest.
However, it is well-known that commis may include unrelated (\ie~tangled) changes~\cite{dias2015untangling}, \eg~it may perform migration and add/remove/update assertions.
To avoid this problem, we manually detect \emph{isolated migrations}, that is, migrations that simply replace unittest by pytest, and no other unrelated changes are involved.
Moreover, to avoid noise caused by large commits, we filter commits that modified more than 5 file tests.

\section{Dataset Description}

Our dataset has been curated from the test suites of 100 highly popular Python software systems.
The 100 selected systems come from our study that empirically analyzed the migration from unittest to pytest~\cite{barbosa2022}.
It includes systems that are broadly adopted worldwide, such as Pandas, Flask, Requests, Cookiecutter, Aiohttp, Ansible, to name a few.\footnote{The complete list of systems can be found in the original dataset: \url{https://doi.org/10.5281/zenodo.5594254}.}

First, we executed the migration detection tool (as described in Section~\ref{subsec:step1}) in the 100 selected projects and detected 690 migration commits in 37 projects.
Of the 690 migration commits, we manually detected 923 isolated migrations (described in Section~\ref{subsec:step2}), which were used to create our dataset.
Thus, TestMigrationsInPy contains 923 real-world migrations from unittest to pytest.

\smallskip

\noindent\textbf{Dataset Structure:}
To facilitate navigation in our dataset, we organized it as a repository in GitHub.
Each project has a list of migration commits.
Each migration commit has a list of isolated migrations and their respective code (located in folder \texttt{diff}) and a migration summary (located in file \texttt{output.info}):

\begin{itemize}

    \item Migration code (\texttt{diff}): contains the migrations of a commit; each migration includes the test code before (with unittest) and after (with pytest) the migration.
    
    \item Migration summary (\texttt{output.info}): contains the commit hash, the number of changed files, the number of migrations in the commit, and type. The ``type'' attribute includes information about the migration type, for example, whether it changes assertions or fixtures (see an example in the next section).
    
\end{itemize}

\section{Dataset Examples}

Migrations from unittest to pytest do not have the same level of difficulty.
For example, migrating assertions may be simple because one only needs to replace the unittest assertions with pytest ones and adapt the data being compared.
In contrast, a migration that replaces unittest setups/teardowns by pytest fixtures may be harder to achieve because there is no direct mapping.
Next, we present two examples to illustrate such scenarios.

\subsection{Example 1: Simple Migration of Assertions}

\begin{figure}[h]
     \centering
         \fbox{\includegraphics[width=0.47\textwidth]{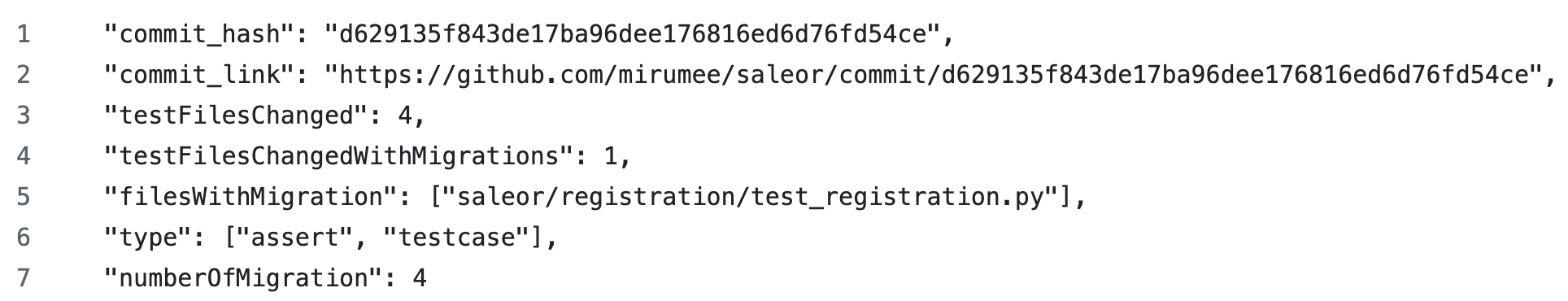}}
         \caption{Example of \texttt{output.info} file (Saleor, commit \#1).}
        \label{fig:saleor_output}
\end{figure}

\begin{figure}[h]
     \centering
     \begin{subfigure}[b]{0.45\textwidth}
         \centering
         \fbox{\includegraphics[width=\textwidth]{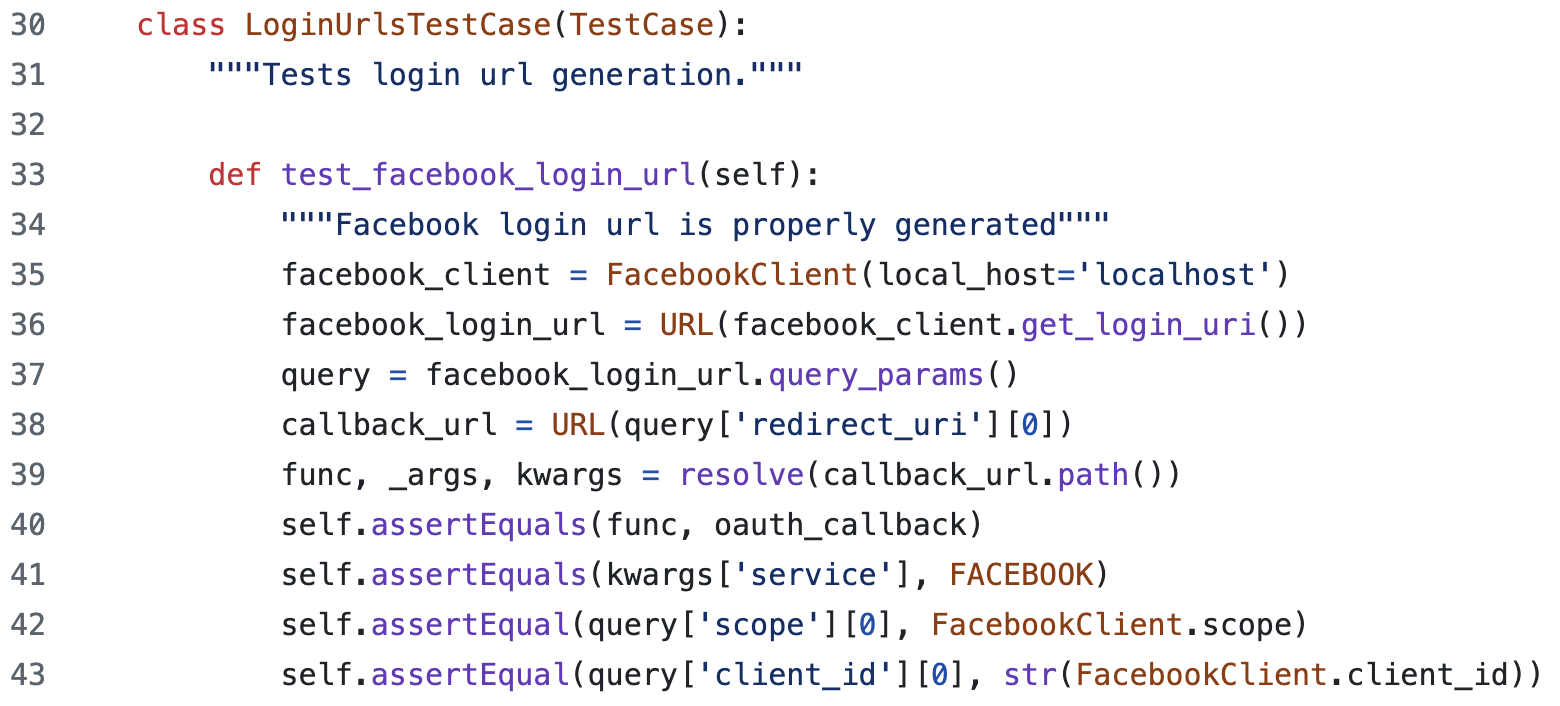}}
         \caption{Test code before the migration (with unittest).}
         \label{fig:ex1a}
     \end{subfigure}
     \begin{subfigure}[b]{0.45\textwidth}
         \centering
         \fbox{\includegraphics[width=\textwidth]{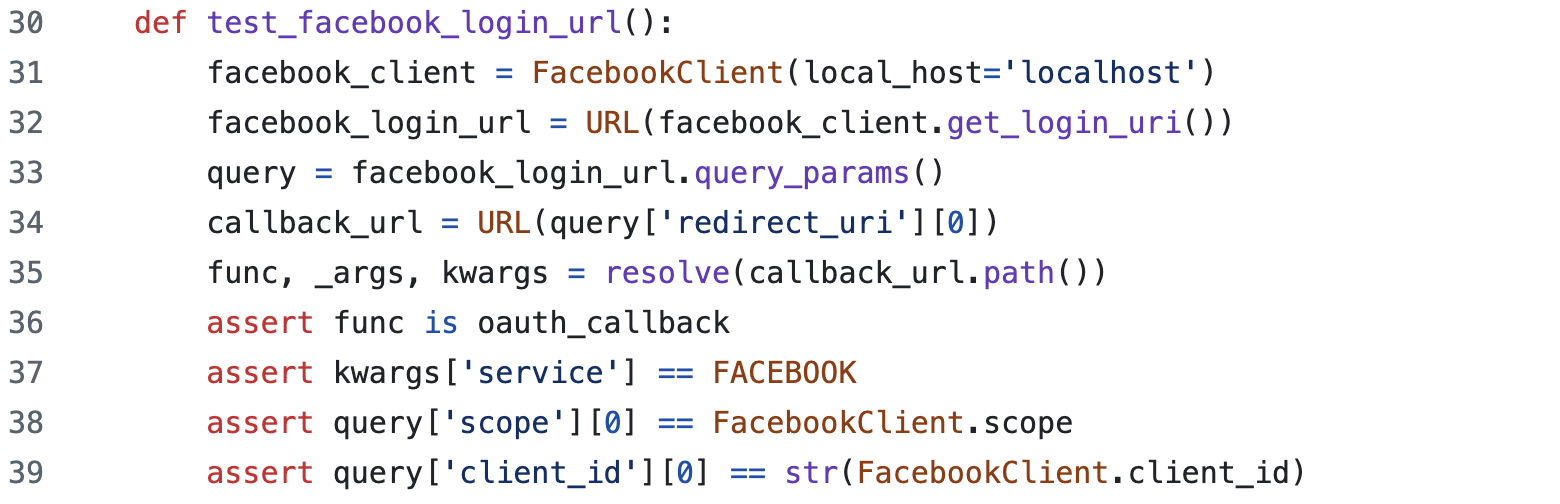}}
         \caption{Test code after the migration (with pytest).}
         \label{fig:ex1b}
     \end{subfigure}
    \caption{Example of migration of asserts (Saleor, commit \#1, migration \#1)}
    \label{fig:saleor}
\end{figure}

Project Saleor has 1 migration commit.\footnote{List of migration commits in Saleor: \url{https://github.com/altinoalvesjunior/TestMigrationsInPy/tree/main/projects/saleor}}
Such migration commit has 4 migrations, as summarized in its \texttt{diff}\footnote{Migration code (Saleor, commit \#1): \url{https://github.com/altinoalvesjunior/TestMigrationsInPy/tree/main/projects/saleor/1/diff}} and detailed in \texttt{output.info}\footnote{Migration summary (Saleor, commit \#1): \url{https://github.com/altinoalvesjunior/TestMigrationsInPy/blob/main/projects/saleor/1/output.info}} (see Figure~\ref{fig:saleor_output}).
Note that the ``type'' attribute informs us that these migrations only involve assertion changes.
For instance, migration \#1 migrates test \texttt{test\_\-face\-book\_\-login\_\-url} from unittest\footnote{Test before: \url{https://github.com/altinoalvesjunior/TestMigrationsInPy/blob/main/projects/saleor/1/diff/mig1-before-test_registration.py}} to pytest,\footnote{Test after: \url{https://github.com/altinoalvesjunior/TestMigrationsInPy/blob/main/projects/saleor/1/diff/mig1-after-test_registration.py}} as detailed in Figure~\ref{fig:saleor}.
In this case, the unittest \texttt{self.assert*} statements are replaced by the pytest \texttt{assert} ones.
Notice that it is needed to adapt the data being compared, for instance, \texttt{self.assertEquals(func, oauth\_callback)} is replaced by \texttt{assert func is oauth\_callback}.
Other changes would be required depending on the assertion being used in unittest.
For instance, \texttt{self.assertIn(a,b)} should be replaced by \texttt{assert a in b}, and \texttt{self.assertIsInstance(a,b)} should be replaced by \texttt{assert isinstance(a, b)}, to name a few.

\subsection{Example 2: Complex Migration of Fixtures and Assertions}

\begin{figure}[h]
     \centering
         \fbox{\includegraphics[width=0.47\textwidth]{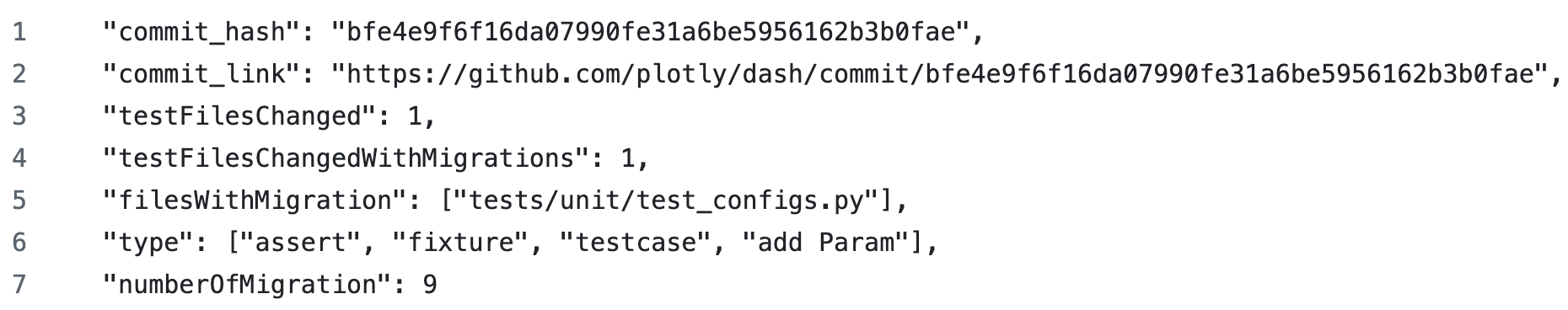}}
         \caption{Example of \texttt{output.info} file (Dash, commit \#1).}
        \label{fig:dash_output}
\end{figure}

\begin{figure}[h]
     \centering
     \begin{subfigure}[b]{0.45\textwidth}
         \centering
         \fbox{\includegraphics[width=\textwidth]{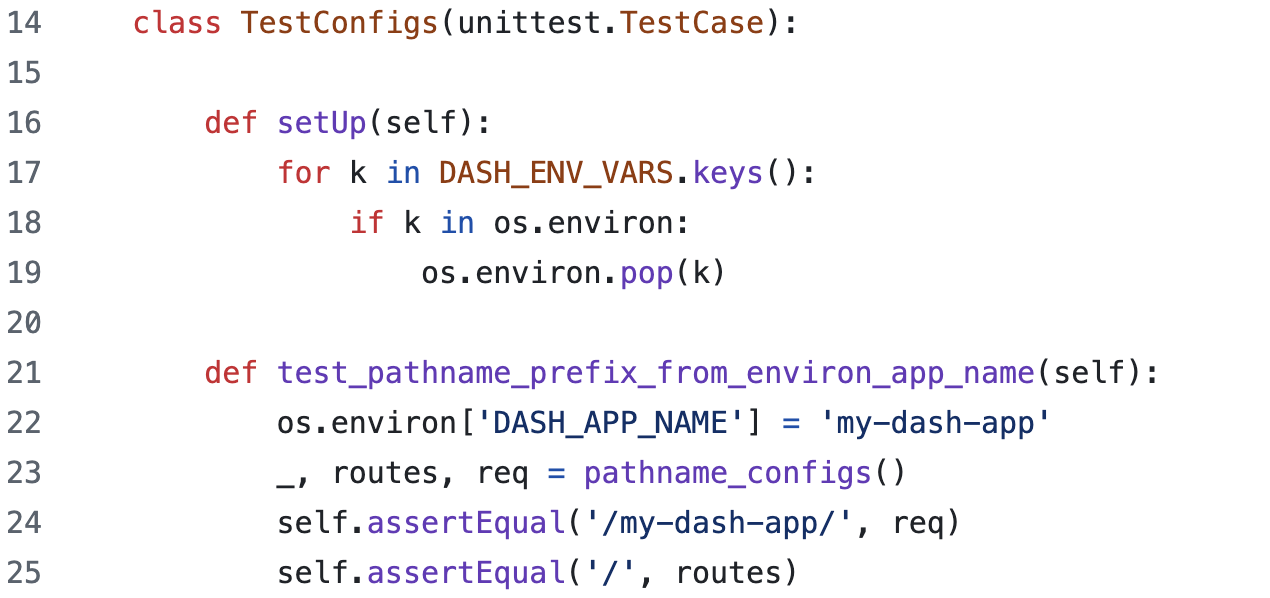}}
         \caption{Test code before the migration (with unittest).}
         \label{fig:ex1a}
     \end{subfigure}
     \begin{subfigure}[b]{0.45\textwidth}
         \centering
         \fbox{\includegraphics[width=\textwidth]{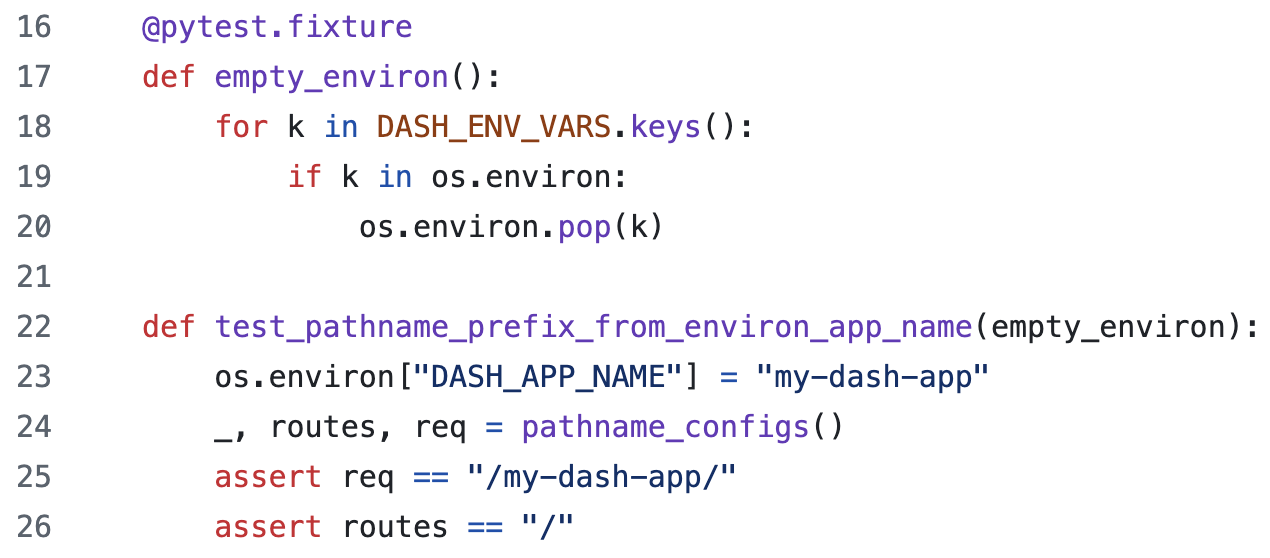}}
         \caption{Test code after the migration (with pytest).}
         \label{fig:ex1b}
     \end{subfigure}
    \caption{Example of migration fixtures and assertions (Dash, commit \#1, migration \#4)}
    \label{fig:dash}
\end{figure}

Project Dash has 4 migration commits.\footnote{List of migration commits in Dash: \url{https://github.com/altinoalvesjunior/TestMigrationsInPy/tree/main/projects/dash}}
Migration commit \#1 has 9 migrations, as presented in its \texttt{diff}\footnote{Migration code (Dash, commit \#1): \url{https://github.com/altinoalvesjunior/TestMigrationsInPy/tree/main/projects/dash/1/diff}} and summarized in its \texttt{output.info}\footnote{Migration summary (Dash, commit \#1): \url{https://github.com/altinoalvesjunior/TestMigrationsInPy/blob/main/projects/dash/1/output.info}} (see Figure~\ref{fig:dash_output}).
Notice that the ``type'' attribute shows that the migration involves both assertion and fixture changes.
As an example, migration \#4 migrates a setup and a test from unittest\footnote{Test before: \url{https://github.com/altinoalvesjunior/TestMigrationsInPy/blob/main/projects/dash/1/diff/mig4-before-test_configs.py}} to pytest,\footnote{Test after: \url{https://github.com/altinoalvesjunior/TestMigrationsInPy/blob/main/projects/dash/1/diff/mig4-after-test_configs.py}} as detailed in Figure~\ref{fig:dash}.
In this case, there are three major changes.
First, the unittest test class \texttt{TestConfigs} (with inheritance to \texttt{TestCase}) is removed, as it is not needed in pytest.
Second, the unittest \texttt{setUp} method is replaced by the pytest fixture \texttt{empty\_environ}, which is annotated with \texttt{@pytest.fixture}.
Third, the unittest test method \texttt{test\_\-path\-name\_\-prefix\_\-from\_\-environ\_\-app\_\-name} is replaced by the pytest test function with the same name.
The new pytest test function receives the fixture \texttt{empty\_environ} as a parameter.
When pytest runs a test, it looks at the parameters in that test function’s signature and then searches for fixtures with the same names as those parameters~\cite{pytest, barbosa2022}.
Once pytest finds them, it runs those fixtures, captures what they returned, and passes those values into the test function as arguments~\cite{pytest, barbosa2022}.

As a more challenging migration of fixtures, in project pyvim, the unittest \texttt{setUp} method is split into four pytest fixture functions: \texttt{prompt\_buffer}, \texttt{editor\_buffer}, \texttt{window}, and \texttt{tab\_page}.\footnote{\url{https://github.com/prompt-toolkit/pyvim/commit/7e1c7bfb505cefba468}}

\section{Dataset Usage: Support the Development of Novel Framework Migration Solutions}

Multiple studies explore migration, both empirically~\cite{barbosa2022, martinez2020and} and by proposing automatic migration solutions~\cite{esem2024_api_migration_llm, dias2015untangling, hora2015apiwave, zhong2010mining, phan2017statistical}.
For example, Barbosa and Hora empirically explored how developers migrate Python tests from unittest to pytest~\cite{barbosa2022}.
In a related line, Martinez and Mateus studied the migration from Java to Kotlin~\cite{martinez2020and, martinez2021did}.
Recently, Large Language Models (LLMs) have been adopted in multiple software engineering tasks~\cite{fan2023large, monteiro2023end, liang2023can, tufano2023predicting, georgsen2023beyond, hou2023large, hora2024predicting, esem2024_api_migration_llm, schafer2023empirical, alshahwan2024automated}, including code migration~\cite{di2025deepmig, esem2024_api_migration_llm}.
Di Rocco~\etal~proposed DeepMig, a transformer-based approach to support coupled library and code migrations in Java~\cite{di2025deepmig}.
Almeida~\etal~provided an initial study to explore automatic library migration using LLMs~\cite{esem2024_api_migration_llm}.
Both LLM-based migrations presented promising results~\cite{di2025deepmig, esem2024_api_migration_llm}.

Our dataset contains real-world migrations performed by developers from the testing framework unittest to pytest.
We manually verified the migrations to reduce the possible noise caused by unrelated, tangled changes~\cite{dias2015untangling}.
Therefore, we foresee the following usages for TestMigrationsInPy.

\begin{boxH}
\textbf{Usage 1:}
Future research proposing novel solutions to migrate frameworks in Python can rely on TestMigrationsInPy as a ground truth.
For example, to explore migration with LLMs, the test code before (with unittest) can be migrated with such models, and the test code after (with pytest) can be used as the ground truth for the LLM-migrated test.
\end{boxH}

Our dataset also includes information about the migration type, such as changes in assertions or fixtures.
As migrations of fixtures are more complex than migrations of assertions, they can be classified according to their difficulty level.

\begin{boxH}
\textbf{Usage 2:}
TestMigrationsInPy enables novel migration solutions to be verified effectively, from simpler assertion migrations to more complex fixture migrations. 
\end{boxH}

In this context, we have used GPT-4o and TestMigrationsInPy to migrate Python tests from unittest to pytest.
Overall, our initial results show that GPT-4o can be used to accelerate the migration process from unittest to pytest.
However, we observed that developers should pay attention to fixing minor wrong updates that the model can perform, particularly when migrating fixtures.


\section{Limitations}

Our dataset has been curated from the test suites of 100 highly popular Python software systems~\cite{barbosa2022}.
Despite these systems being relevant and real-world, they do not represent the entire Python ecosystem.
Particularly, less popular projects may adopt migration practices that are uncommon in more widely-used projects.
Further versions of the dataset can include migrations from more projects, allowing a more comprehensive understanding of the migration landscape.

\section{Related Work}

\subsection{Code Migration}

Framework evolution and migration are research topics largely explored by the literature in multiple ecosystems~\cite{barbosa2022, lamothe2021systematic, wang2020exploring, xavier2017historical, li2018characterising, brito2020you, hora2015developers, brito2018use, robbes2012developers, li2013does, sawant2019react, sawant2018features, nascimento2021javascript, malloy2019empirical}.
In the context of testing framework migration, Barbosa and Hora empirically explored how developers migrate Python tests from unittest to pytest~\cite{barbosa2022}.
In many cases, the migration was not simple, taking a long period to conclude or never concluded at all.
In a related research line, Martinez and Mateus studied the migration from Java to Kotlin~\cite{martinez2020and, martinez2021did}.
Kotlin is interoperable with Java, thus, developers can migrate gradually.
Overall, the migration occurred to access features only available in Kotlin and to obtain safer code. 

Recently, LLMs have been adopted in multiple software engineering tasks, including generating tests, refactoring, fixing bugs, and supporting code review~\cite{fan2023large, monteiro2023end, liang2023can, tufano2023predicting, georgsen2023beyond, hou2023large, hora2024predicting, esem2024_api_migration_llm, di2025deepmig, schafer2023empirical, alshahwan2024automated}.
Notably, it has demonstrated significant results in code generation~\cite{openai2023gpt4, fan2023large}.
In this context, Di Rocco~\etal~proposed DeepMig, a transformer-based approach to support coupled library and code migrations in Java~\cite{di2025deepmig}.
The research presents promising results, showing that DeepMig can recommend both libraries and code; in several projects with a perfect match.
Almeida~\etal~provided an initial study to explore automatic library migration using LLMs~\cite{esem2024_api_migration_llm}.
With GPT-4o, the authors migrated a client application to a newer version of SQLAlchemy, a Python ORM library.
TestMigrationsInPy can support the development of novel solutions to migrate frameworks in Python.

\subsection{Datasets to Support Testing Research}

Multiple datasets have been proposed to support software testing research~\cite{hora2024testdossier, tufano2022methods2test, kicsi2020testroutes, gao2018jbench, bui2023snapshot}.
For example, TestDossier is a dataset of tested values automatically extracted from the execution of Python tests~\cite{hora2024testdossier}.
Methods2Test is a dataset of focal methods mapped to test cases extracted from Java projects~\cite{tufano2022methods2test}.
TestRoutes is a test-to-code dataset containing traceability information for Java test cases ~\cite{kicsi2020testroutes}.
Jbench is a dataset of data races for concurrency testing~\cite{gao2018jbench}.
TestMigrationsInPy contributes to the software testing literature with a novel dataset to support test migration research.

\section{Conclusion}

We proposed TestMigrationsInPy, a dataset of test migrations from unittest to pytest.
TestMigrationsInPy contains real-world migrations performed by developers.

\noindent\textbf{Further Improvements:}
First, our dataset can include less popular projects to have a better overview of the migration landscape.
Second, the migrations can have a more specific classification for the unittest test code, for instance, with specific used assertions and fixtures (\eg~\texttt{setUp}, \texttt{setUpClass}, \texttt{tearDown}, etc.).
Third, the pytest test code can have a more specific classification of the pytest features used in the migrated test (\eg~parametrized tests).

\section*{Acknowledgments}
This research is supported by CAPES, CNPq, and FAPEMIG.

\bibliographystyle{IEEEtran}
\bibliography{main}

\end{document}